\documentclass[10pt,conference,table]{IEEEtran}

\usepackage[utf8]{inputenc}
\usepackage{amsmath}
\usepackage{amssymb}
\usepackage{graphicx}
\usepackage{siunitx}
\usepackage{array}
\usepackage{listings}
\usepackage[frozencache]{minted}
\usepackage{soulutf8}
\usepackage{tikz}
\usetikzlibrary{decorations.pathreplacing}
\usepackage{makecell}
\usepackage{float}
\usepackage{tabularx}
\newcolumntype{Y}{>{\centering\arraybackslash}X}
\usepackage{balance}
\usepackage{subcaption}
\usepackage{hyperref}
\usepackage{url}
\usepackage{xspace}
\usepackage{tcolorbox}
\usepackage{enumitem}
\usepackage{multirow}
\usetikzlibrary{shapes}
\usetikzlibrary{patterns, fit}
\usepackage{adjustbox}
\usepackage{pifont}
\setminted[java]{
xleftmargin=15pt,
framesep=2mm,
linenos=true,
breaklines=true,
tabsize=2,
encoding=utf8,
frame=none,
fontsize=\scriptsize,
escapeinside=|| 
}

\newcommand{\highlight}[1]{\begin{tcolorbox}[leftrule=0mm,rightrule=0mm,toprule=0mm,bottomrule=0mm,left=0pt,right=0pt,top=0pt,bottom=0pt]
#1
\end{tcolorbox}
}

\makeatletter
\newcommand{\ostar}{\mathbin{\mathpalette\make@circled\star}}
\newcommand{\make@circled}[2]{%
  \ooalign{$\m@th#1\smallbigcirc{#1}$\cr\hidewidth$\m@th#1#2$\hidewidth\cr}%
}
\newcommand{\smallbigcirc}[1]{%
  \vcenter{\hbox{\scalebox{0.97778}{$\m@th#1\bigcirc$}}}%
}
\newbox\dottedarrow@box
\setbox\dottedarrow@box\hbox
  {%
    \begin{tikzpicture}
      \draw[dotted,->] (0,0) -- (1.5em,0);
    \end{tikzpicture}%
  }
\newcommand*\dottedarrow
  {\relax\ifmmode\expandafter\dottedarrow@m\else\expandafter\dottedarrow@t\fi}
\newcommand*\dottedarrow@t[1][1.5em]
  {\resizebox{#1}{!}{\raisebox{.5ex}{\usebox\dottedarrow@box}}}
\newcommand*\dottedarrow@m[1][]
  {%
    \if\relax\detokenize{#1}\relax
      \mathchoice
        {\dottedarrow@t}
        {\dottedarrow@t}
        {\dottedarrow@t[1.1em]}
        {\dottedarrow@t[0.9em]}%
    \else
      \dottedarrow@t[#1]%
    \fi
  }
\makeatother

\newcolumntype{g}{>{\columncolor{gray!30}}l}
\newcolumntype{h}{>{\columncolor{gray!30}}c}

\newcommand{\dcircle}[1]{\ding{\numexpr171 + #1}}

\newcommand{\tool}[0]{\textsc{CoDoC}\xspace}
\newcommand{\flowdroid}[0]{\textsc{FlowDroid}\xspace}

\newcommand{\gp}[0]{\textsc{Google Play}\xspace}

\newcommand{\susi}[0]{\textsc{SuSi}\xspace}
\newcommand{\source}[0]{\texttt{SOURCE}\xspace}
\newcommand{\sink}[0]{\texttt{SINK}\xspace}
\newcommand{\neither}[0]{\texttt{NEITHER}\xspace}
\newcommand{\ctov}[0]{\textsc{Code2Vec}\xspace}
\newcommand{\sbert}[0]{\textsc{Sentence-BERT}\xspace}
\newcommand{\bert}[0]{\textsc{BERT}\xspace}

\newcommand{\susiNew}[0]{\textsc{SuSi\_new}\xspace}
\newcommand{\susiOriginal}[0]{\textsc{SuSi\_Original}\xspace}

\newcommand{\numSources}[0]{231\xspace}
\newcommand{\numSinks}[0]{130\xspace}
\newcommand{\numNeither}[0]{654\xspace}
\newcommand{\total}[0]{1015\xspace}

\def\BibTeX{{\rm B\kern-.05em{\sc i\kern-.025em b}\kern-.08em
    T\kern-.1667em\lower.7ex\hbox{E}\kern-.125emX}}
    
\begin{document}

\title{Negative Results of Fusing Code and Documentation for Learning to Accurately Identify Sensitive Source and Sink Methods\\
\LARGE An Application to the Android Framework for Data Leak Detection
}

\author{\IEEEauthorblockN{Jordan Samhi\IEEEauthorrefmark{1}, Maria Kober\IEEEauthorrefmark{3}\textsuperscript{\textsection}, Abdoul Kader Kabore\IEEEauthorrefmark{1}, Steven Arzt\IEEEauthorrefmark{2}, Tegawendé F. Bissyandé\IEEEauthorrefmark{1},
Jacques Klein\IEEEauthorrefmark{1}}
\IEEEauthorblockA{\IEEEauthorrefmark{1} SnT,
University of Luxembourg, firstname.lastname@uni.lu}
\IEEEauthorblockA{\IEEEauthorrefmark{2} Fraunhofer Institute for Secure Information Technology, Darmstadt, Hessen, Germany, steven.arzt@sit.fraunhofer.de}
\IEEEauthorblockA{\IEEEauthorrefmark{3} mariakober.research@gmx.com}}

\pagestyle{plain}

\maketitle
\begingroup\renewcommand\thefootnote{\textsection}
\footnotetext{Most of the work was completed while Maria Kober was present at Fraunhofer SIT}
\endgroup

\begin{abstract}
Apps on mobile phones manipulate all sorts of data, including sensitive data, leading to privacy-related concerns.
Recent regulations like the European GDPR provide rules for the processing of personal and sensitive data, like that no such data may be leaked without the consent of the user.

Researchers have proposed sophisticated approaches to track sensitive data within mobile apps, all of which rely on specific lists of sensitive \source and \sink API methods. The data flow analysis results greatly depend on these lists' quality. Previous approaches either used incomplete hand-written lists that quickly became outdated or relied on machine learning. The latter, however, leads to numerous false positives, as we show.

This paper introduces \tool, a tool that aims to revive the machine-learning approach to precisely identify privacy-related \source and \sink API methods. In contrast to previous approaches, \tool uses deep learning techniques and combines the source code with the documentation of API methods.
Firstly, we propose novel definitions that clarify the concepts of sensitive \source and \sink methods.
Secondly, based on these definitions, we build a new ground truth of Android methods representing sensitive \source, \sink, and \neither (i.e., no source or sink) methods that will be used to train our classifier.

We evaluate \tool and show that, on our validation dataset, it achieves a precision, recall, and F$_1$ score of 91\% in 10-fold cross-validation, outperforming the state-of-the-art \susi when used on the same dataset.
However, similarly to existing tools, we show that in the wild, i.e., with unseen data, \tool performs poorly and generates many false positive results. Our findings, together with time-tested results of previous approaches, suggest that machine-learning models for abstract concepts such as privacy fail in practice despite good lab results.
To encourage future research, we release all our artifacts to the community.

\end{abstract}

\section{Introduction}
\label{sec:introduction}

Given the ubiquity of mobile devices nowadays and the proliferation of apps installed and used by end users, Android apps' analysis has become a common topic in software engineering research.
Numerous approaches have been proposed to check security properties, detect malicious code, and detect program bugs. These approaches leverage techniques such as dynamic analysis~\cite{van2013dynamic, 10.1145/2592791.2592796, enck2014taintdroid}, static analysis~\cite{LI201767,10.1109/TDSC.2021.3108057,10.1007/978-3-319-68690-5_12,10.1145/2660267.2660357,7194563,9402001,arzt2014flowdroid,gordon2015information,10.1145/3510003.3510135}, or both (i.e., hybrid analyses)~\cite{brumley2008automatically, bello2018ares}.

The previously mentioned analysis approaches usually consist of several specific techniques that are applied to apps. 
One of them is taint analysis, which 
checks whether data obtained from a given \source method (or any kind of data derived from it, e.g., after some computation) is passed to a \sink method. In the context of Android-based privacy research, a \source is an API method that provides privacy-sensitive data. A \sink is an API method that writes data to the outside, e.g., via the network. 

The need for sensitive \source and \sink lists is ubiquitous in taint analysis.
Indeed, the literature is full of approaches and techniques that set up privacy strategies building on sensitive API methods that allow retrieving sensitive data and/or API methods allowing to expose this kind of data.
The range of approaches relying on sensitive \source and \sink methods that could benefit from a complete and precise list of \source and \sink methods is large: \emph{sensitive data leak detection}~\cite{LI201767,10.1007/978-3-642-30921-2_17,enck2014taintdroid,arzt2014flowdroid,kim2012scandal,10.1007/s10664-021-09943-x,10.1145/3510003.3512766}, \emph{Android component leak detection}~\cite{7272932}, \emph{dynamic policy enforcement}~\cite{rasthofer2014droidforce}, \emph{malware detection}~\cite{8005492,7976989,rathi2018droidmark}, \emph{hidden behavior detection}~\cite{pan2017dark,zhao2020automatic}, \emph{inter-app communication analysis}~\cite{10.1145/1999995.2000018,10.1007/978-3-319-18467-8_34}, \emph{component hijacking vulnerabilities detection}~\cite{10.1145/2382196.2382223,zhang2014appsealer}, the \emph{uncovering of run-time sensitive values}~\cite{rasthofer2016harvesting}, as well as \emph{GDPR compliance checks}~\cite{ferrara2018static,9251060}.

In its public-facing API -- i.e., methods contained in the official online documentation for Android --,
Android 11\footnote{API level 30, which was released in September 2020} contains more than \num{34000} methods. 
However, it has been shown~\cite{7816486} that developers have access to many more methods inside the Android framework that are not directly available in the public-facing API (e.g., hidden to developers using the annotation "@hide").
In Android 11, for example, more than \num{210000} methods are available to developers in total, e.g., through the use of reflection\footnote{Note that reflection can also be used to make private methods accessible since Android does not provide a Java security manager.}.
These large numbers render a manual classification of sources and sinks infeasible. 
Furthermore, additional methods are added in every new release. New frameworks, such as Android Auto or Chromecast, bring in new methods, and thus potentially new sources and sinks as well. Therefore, automatic approaches for identifying sources and sinks are needed.
Several approaches have been proposed in the literature to solve this problem~\cite{10.1007/978-3-642-30921-2_17,nan2018finding,enck2014taintdroid}. \susi~\cite{arzt2013susi}, which is based on supervised machine-learning, is currently one of the most popular approaches and the state-of-the-art~\cite{arzt2014flowdroid,wu2016effective,pan2017dark}. To the best of our knowledge, it is also the most comprehensive and state-of-the-art approach for deriving lists of sources and sinks from frameworks like Android. 

However, the sources list generated by \susi is neither precise, nor specific for privacy analysis. As we show in Section~\ref{sec:motivation}, \susi classifies methods as sources, even though they are clearly irrelevant for privacy analysis. Secondly, \susi relies on technical categories (network information, unique identifiers, etc.) to structure its output. Selecting all categories that could be relevant for privacy analysis leads to a large number of irrelevant APIs being selected as well. These observations are not surprising because \susi makes no upfront assumptions on the sensitivity of the sources yielded.

Further, some methods are misclassified entirely by \susi, e.g., the method \texttt{getScrollIndicatorBounds} of the \texttt{android.view.ViewGroup} class is categorized as a source in the "SMS\_MMS" category. As we show  in Section~\ref{sec:eval:rq5}, these issues lead to a profusion of false positives for any data flow tracker that relies on \susi's source/sink lists.
Several works in the literature~\cite{8776652,JUNAID201692,8952502} have come to similar conclusions. Luo et al. show that \susi's sources list leads to a false positive rate of almost 80\% while trying to detect sensitive data leaks~\cite{8952502}. Further, in accordance with our own findings, they state the following regarding the sources yielded by \susi:

\begin{quote}
\textit{the root cause of these false positives is that their sources [...] are actually inappropriate, i.e., they do not return \textbf{sensitive data}.}
\end{quote}

We note that \susi's evaluation in the original paper~\cite{arzt2013susi} does not highlight these issues, and that \susi performs well on its training data and select examples. However, in the real-world, the false positive rate is much higher.

To the best of our knowledge, no other approach has tackled the problem of automatically identifying sources and sinks in Android since the release of \susi in 2014. In other words, \susi is still the most relevant approach despite its shortcomings. Since the problem remains highly relevant and unsolved regarding sensitive data, we attempted to improve the \susi approach. The \susi features for the supervised machine learning rely entirely on static code analysis on the Android bytecode implementation (the Android platform JAR file). \susi considers individual properties as features, such as method names, parameter types, or method/class modifiers which do not capture the entire semantic of the code.

Our approach, that we named \tool, on the other hand, captures the entire semantics of a given method by taking its' complete source code into account. Additionally, \tool also considers the JavaDoc documentation of the Android API. We observed that the Android documentation, which is fairly extensive for most classes and methods, provides enough guidance to the developer to correctly use the API. We then assumed that analyzing this documentation would also help in more precisely discovering sensitive \source and \sink methods in the Android API.

Lastly, \tool is an attempt to incorporate the ground-breaking advances that have been made in text and code embedding~\cite{10.1145/3290353,reimers2019sentence} and, thus, machine learning since 2014 when \susi was published.

While our evaluation shows that \tool outperforms \susi in the lab, \tool's real-world performance, likewise, is still lacking. Manually inspecting the \source and \sink methods identified in a set of previously unseen API methods from the Android framework reveals many false positives. We, therefore, argue that even adding documentation and improving the machine learning techniques do not solve the problem of accurately identifying privacy-related sources and sinks in Android. Even with more and better training data and careful optimization of the training, the overall goal remains elusive. We argue that the semantic gap between an individual API method (code or documentation) and an abstract concept such as user privacy are unlikely to be closed by supervised machine learning. Instead, novel approaches are necessary.

Further, we call for a more careful evaluation of machine learning results. 
In the lab studies based on 10-fold cross-validation, \susi is sufficient, and \tool is even better. Still, on real-world data, i.e., previously unseen methods from the Android framework, both fail to meet expectations.

Overall, we make the following contributions:
\begin{itemize}
    \item we propose \tool: a novel, fully-automated, deep-learning-based approach to detect sensitive \source and \sink methods in the Android framework based on API method source code and documentation;
    \item we release a new ground-truth of methods labeled as sensitive \source, \sink, or \neither;
    \item we evaluate \tool and show that it outperforms the state-of-the-art \susi on a small evaluation dataset, reaching a precision, recall, and F$_1$ score of 91\% in the lab; 
    \item we apply \tool on public methods from the Android framework and show that, likewise \susi, it yields a high rate of false positives;
    \item We release our open-source prototype \tool to the community and all the artifacts used in our study at:
    \begin{center}
       \url{https://github.com/JordanSamhi/CoDoC}
    \end{center}
\end{itemize}

\section{Background}
\label{sec:background}

In this section, we provide the reader with context for the work presented in this paper. 

\subsection{Taint Analysis}
\label{sec:taint_analysis}

Taint analysis is a particular dataflow analysis that tracks data through the control flow graph of a program. If a variable $V$ is assigned the return value of a specific function, like a \source method, it becomes tainted. If a tainted value is assigned to some other variable $W$, this variable $W$ gets tainted as well. The same applies if $W$ is assigned the result of some operation on an already tainted variable $V$. In other words: the taint is propagated. When a tainted variable is passed to a \sink function as a parameter, a leak is reported, as the value derived from the \source reached a \sink. In the case of data leak detection in the context of privacy analysis, an example of a \source in the Android Framework is \texttt{getImei()} and an example of a \sink is \texttt{sendTextMessage()}.

\subsection{Text Embedding}
\label{sec:text_embedding}
Our work relies on methods' source code and documentation to train a machine learning model and infer \source and \sink methods. In order to be processed by machine learning algorithms, these textual representations need to be transformed (\emph{embedded}) into numerical representations, i.e., numerical vectors.
In this section, we briefly describe two state-of-the-art techniques for 
this transformation,
namely Sentence-BERT~\cite{reimers2019sentence} and Code2Vec~\cite{10.1145/3290353}. 

\subsubsection{Sentence-BERT} 
\label{sec:sbert}

Method documentation embedding requires efficient natural language processing techniques.
\sbert is a modified and more computationally efficient version of the well-known \bert neural network~\cite{devlin2018bert}. It relies on siamese and triplet network structures to obtain meaningful sentence embeddings.

\subsubsection{\ctov}
\label{sec:code2vec}

Similarly to natural language embedding, making predictions from source code requires code embedding to have a homogeneous representation of different source code inputs.
\ctov embeds Java methods to predict method names.
Methods are transformed into ASTs (Abstract Syntax Trees) to construct path representations between different leaf nodes.
Then, using the attention mechanism~\cite{bahdanau2014neural}, bag of path-contexts are aggregated into a single vector that represents the method body.

\section{Definitions}
\label{sec:definitions}

In the literature, there is no consensus on the definitions of \emph{sensitive} \source and \sink methods, which leads to a lack of clarity in papers related to taint analysis. As described in Section~\ref{sec:taint_analysis}, taint analysis tracks the flow of data from a given \source to a given \sink, no matter the type of data.
However, in most of the papers, the authors mix \emph{sensitive} \source with \source, which makes taint analysis appear as tracking sensitive data, which is not always the case.
Tracking sensitive data is an instance of the more general task of tracking data.

To cope with this problem and provide state-of-the-art approaches that aim at tracking sensitive data with clear terms, we propose the following definitions:

\noindent
\textbf{Definition 1} (Data). 
Any value or reference to a value.

\noindent
\textbf{Definition 2} (Composite Data).
Any structure composed of multiple Data (e.g., an object).

\noindent
\textbf{Definition 3} (Sensitive Data).
Any data or composite data that holds private value(s) that:
\begin{itemize}
    \item can identify users, i.e., usernames and personally identifying data like email address or name
    \item can identify the device, i.e., unique device identifiers
    \item are related data to personal information (of the phone user), e.g., photographs and files, phone calls, SMS
    \item represent data owned by users holding information about other users, e.g., contacts and phone lists, emails, etc.
    \item represent environment and sensor information, including geolocation data, camera, and microphone. 
\end{itemize}

\noindent
\textbf{Definition 4} (Sensitive \source). A function that returns a Sensitive Data. Note that functions that return constant values are never sensitive sources.

\noindent
\textbf{Definition 5} (\sink). A function that sends out one or more Data, Composite Data, or values derived from a Data or a Composite Data from the application memory space. There is no notion of sensitivity for sinks. The nature of the data (more precisely: the \source from which the data was originally obtained) passed to the sink determines whether a leak of sensitive data occurs.
\section{Motivation}
\label{sec:motivation}

Tracking sensitive data within Android apps is of high interest since it is used in numerous security-related approaches~\cite{yang2013appintent,sarwar2013effectiveness,6394931,arzt2014flowdroid,LI201767} and part of legal compliance, e.g., according to the GDPR.
Therefore, there is a need to provide analysts and researchers with sources and sinks lists that precisely enclose sensitive data (cf. Definition 6).

\begin{figure}[ht!]
    \centering
    \includegraphics[scale=.4]{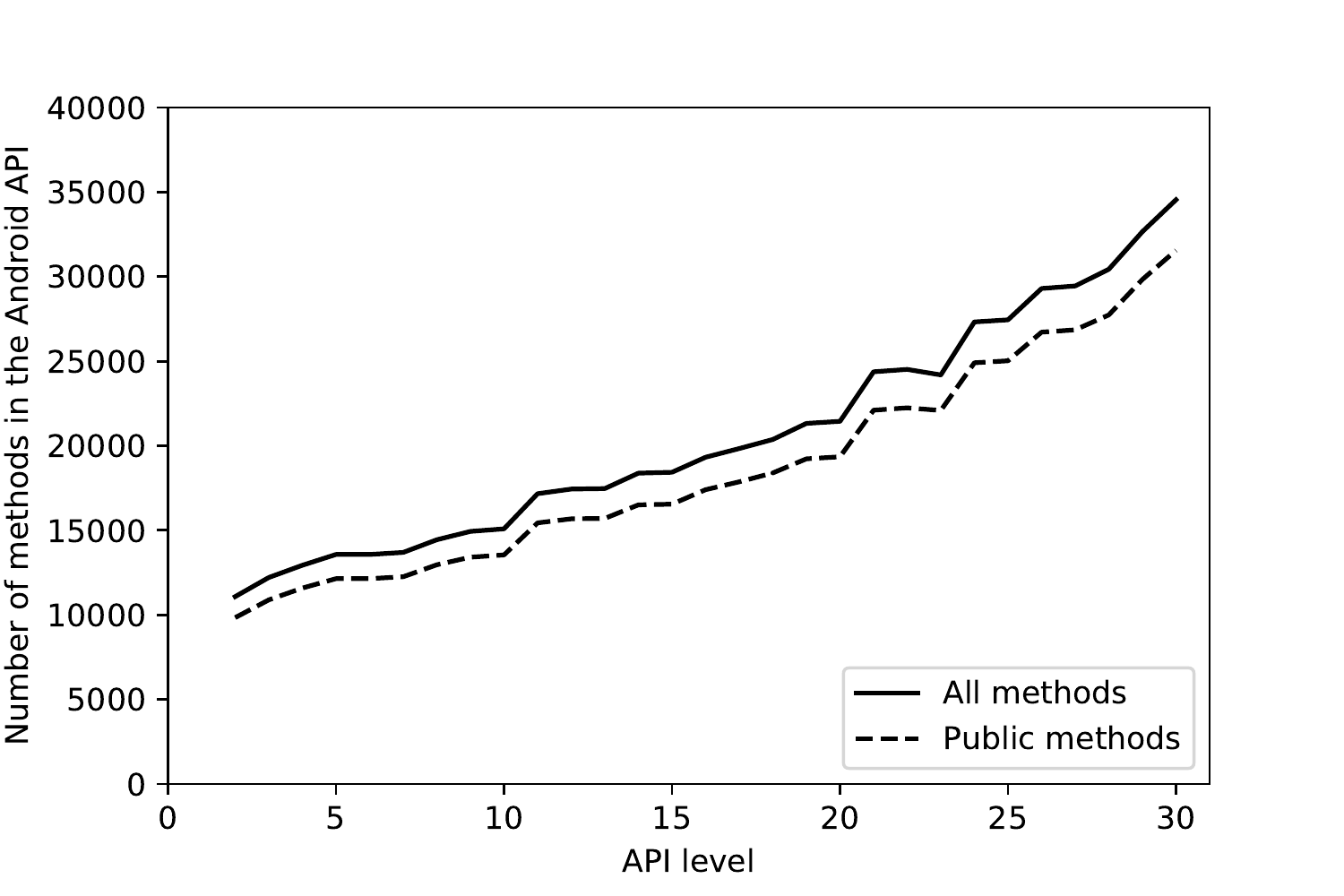}
    \caption{Number of methods in the public-facing Android API by API level}
    \label{fig:api_levels}
\end{figure}

\textbf{Android API:} As briefly explained in Section~\ref{sec:introduction}, the number of public and documented API methods intended for use by Android developers amounts to tens of thousands and increases with every new API version (see Figure~\ref{fig:api_levels}). Still, even identifying all sources and sinks in these documented public API methods is not sufficient, as developers also call methods not intended for direct use, yet present in the Android framework~\cite{7816486}. 
With tens of thousands of public API methods and hundreds of thousands of overall methods, manual classification for every new release is obviously infeasible.

Therefore, automated solutions are needed to produce sensitive sources and sinks lists for every new release. In the following, we explain why the existing state of the art is inappropriate for this task.

\begin{figure}[ht!]
    \centering
    \includegraphics[scale=.4]{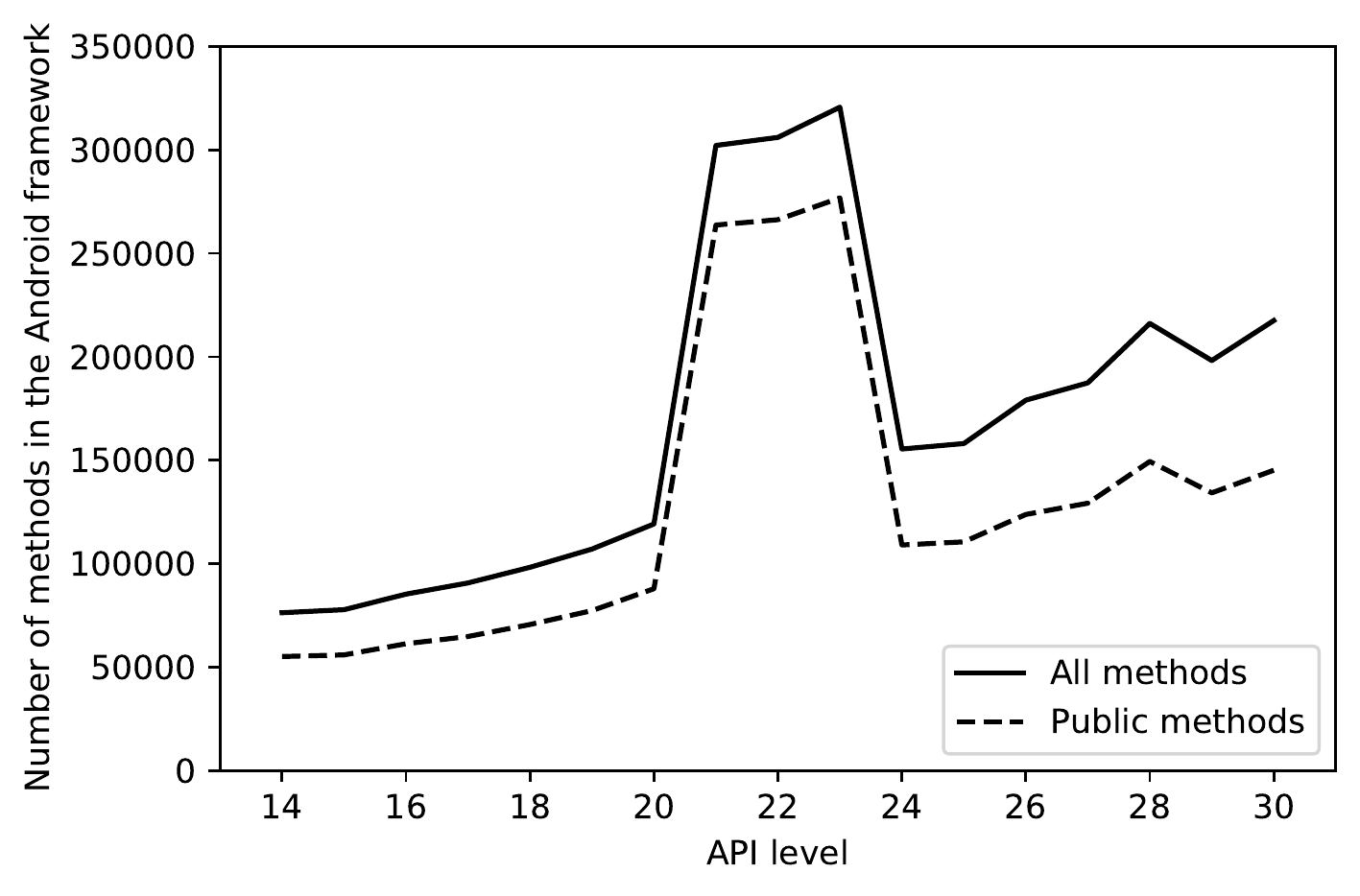}
    \caption{Number of methods in the entire Android framework code by API level}
    \label{fig:api_levels_source_code}
\end{figure}

\textbf{Problem with the existing state of the art:} The state-of-the-art approach \susi~\cite{arzt2013susi} uses machine learning to automatically classify Android \source and \sink methods.
However, it has been shown several times~\cite{8776652,JUNAID201692,8952502} that \susi's lists are inappropriate since it is not specific to a particular analysis like tracking sensitive data. Therefore it produces many false positives and forces analysts to manually select appropriate \source and \sink methods.

Consider the example in Listing~\ref{code:motivation}.
In line 4, a \texttt{Rational} object is created from two integers. In line 5, the integer representation of the \texttt{Rational} is retrieved using the method \texttt{intValue()} and stored in variable \texttt{value}.
Eventually, this \texttt{value} is sent out of the device via SMS. Since \susi wrongly considers the method \texttt{intValue()} as a \source and the method  \texttt{sendTextMessage()} as a \sink, a taint analysis based on \susi will report a leak. This leak, however, is irrelevant in the context of security and privacy as the \source is not sensitive. Thus, analysts will consider it a false positive when aiming to detect sensitive data leaks in Android apps.

We aim to  improve upon the state of the art by \emph{producing a more adequate ground truth to train an improved machine learning model based on method documentation and source code}, unexplored until now. 

\begin{listing}
\begin{minted}{java}
public void method() {
    int p = 7;
    int q = 4;
    Rational r = new Rational(p, q);
    int value = r.intValue();
    SmsManager s = SmsManager.getDefault();
    s.sendTextMessage("0", null, value, null, null);
}
\end{minted}
    \caption{Example of \susi non-sensitive data leak}
    \label{code:motivation}
\end{listing}
\section{Approach}
\label{sec:approach}

\begin{figure*}
    \centering
    \begin{adjustbox}{width=.8\linewidth,center}
    \input{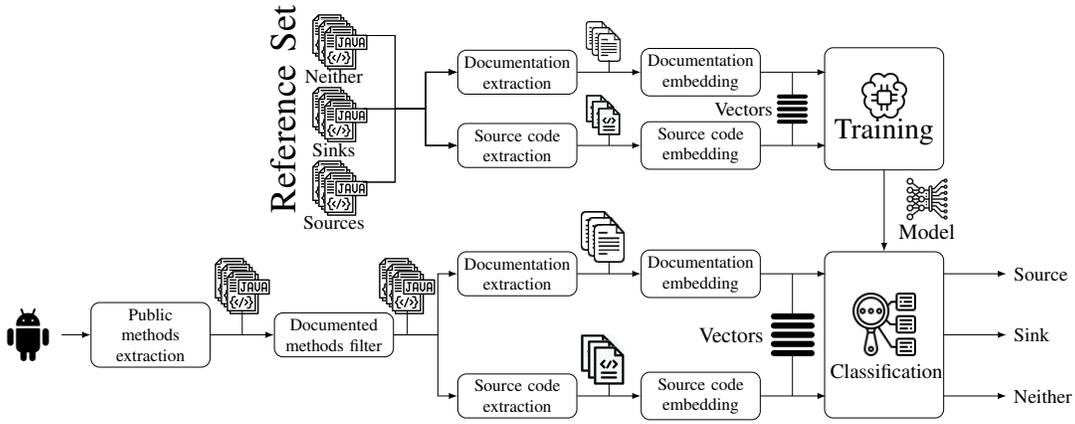}
    \end{adjustbox}
    \caption{Overview of the \tool approach}
    \label{fig:overview}
\end{figure*}

In this paper, we aim to automatically identify sensitive \source and \sink methods in the Android framework among all API methods available to developers (i.e., $>$ \num{210000} in Android 11) using supervised machine learning. Figure~\ref{fig:overview} shows an overview of our approach.

Similar to \susi, we build our training data by manually labeling Android methods. We consider a method as a sensitive \source if it matches Definition 4\footnote{in the following, when we refer to \source, we mean "Sensitive" \source}, a \sink if it matches Definition 5, and \neither otherwise. In contrast to \susi, our approach then uses features extracted from the code as well as the documentation to train a machine-learning model on our ground truth. This is a key difference to \susi, which only uses distinct properties extracted from parts of the code, such as method names and parameters or class and method modifiers. Further, \susi completely disregards the method's documentation, which \tool includes.
Moreover, as we rely on the entire source code of a method, we are able to capture the entire semantic of it. 

We finally use our generated model to predict new sensitive \source and \sink methods from the Android framework methods.
We explain the individual steps in the following sections. First, we give details about our manual labeling of Android methods in Section~\ref{sec:manual_labeling}. Then, in Section~\ref{sec:data_engineering}, we explain what features were chosen for training our models. Lastly, in Section~\ref{sec:architecture}, we explicit on what machine learning models our approach builds upon.

\subsection{Manual Labeling}
\label{sec:manual_labeling}

Since our approach relies on supervised machine learning algorithms, labeled data is needed to train our model.
However, manual labeling is a challenging and time-consuming task, especially if we randomly chose methods from the Android framework to label one by one. Further, finding a \source or a \sink through random picking is highly unlikely as most methods in the Android framework are neither.
Therefore, we opted for a better strategy divided into three phases:

\noindent
\textbf{Phase 1:}
The authors first constituted a \emph{golden dataset} based on well-known methods that return sensitive data described in the literature~\cite{7546513,arzt2014flowdroid,enck2014taintdroid,pan2017dark,enck2011study,10.1145/2660267.2660359}.
These methods span across classes such as: \texttt{TelephonyManager}, \texttt{AccountManager}, \texttt{LocationManager}, \texttt{Sms\-Message}, or \texttt{SensorManager}.
This step yielded an initial set of 39 \source and 35 \sink methods.

\noindent
\textbf{Phase 2:} As explained in Section~\ref{sec:motivation}, \susi can generate lists of sources and sinks (from its own definition, i.e., not restricted to sensitive methods). We applied \susi on Android API version 30 to generate additional pre-selected input that we hand-labeled as training data for \tool. As described previously, randomly picking methods from the Android API would mostly lead to methods that are neither sources nor sinks. Therefore, we opted to focus hand-labeling on methods that are more likely to be relevant.
We concatenated the list of sources and the list of sinks computed by \susi to obtain a full list of methods $M$ that \susi considers relevant. Note that we did not manually post-process methods that \susi classified as neither a source or a sink. Two of the authors then applied manual post-processing as follows. 

One author started from the top of each list manually classify each method in the respective list. The other author started from the bottom of each list with same task. For each method $m \in M$, the authors independently read the documentation and the source code to be able to classify it as a \texttt{SOURCE}, a \texttt{SINK}, or \texttt{NEITHER} based on the definitions described in Section~\ref{sec:definitions}.

This step leads to three lists per author: a \dcircle{1} \source list; a \dcircle{2} \sink list; and \dcircle{3} a \neither list.

\noindent
\textbf{Phase 3:}
The third phase aimed at calculating the inter-rater agreement between the data labeled by both authors in phase 2. We use inter-author agreement as a quality measure for our hand-labeled dataset, which is later used for training the \tool classifier.
To do so, both authors together alternately verified the results of each other and noted the agreement.
Eventually, a Cohen's Kappa coefficient~\cite{doi:10.1177/001316446002000104} was computed to evaluate the level of agreement.
Due to the clear definitions given in Section~\ref{sec:definitions}, both authors reached a perfect agreement level of 1. In phase 2, both authors classified 192 methods as sources, resulting in a total of \numSources \source methods from phase 1 and phase 2; 95 methods as sinks, resulting in a total of \numSinks \sink methods; as well as \numNeither \neither methods. In total, we have a set of \total API methods for model training.

\subsection{Data collection and representation}
\label{sec:data_engineering}

Our approach relies on two different types of input: \dcircle{1} the documentation of a method; and \dcircle{2} the source code of a method.
This section explains how these data were gathered and transformed into numerical value vectors.

\noindent
\textbf{Data Collection:} 
As an open-source project, the Android source code is directly available on the Internet~\footnote{\url{https://android.googlesource.com/}}.
We downloaded and parsed it using \textsc{JavaParser}~\cite{javaparser} to extract \emph{public} methods that were \emph{documented}, and \emph{implemented} (i.e., concrete methods).
A method that is documented means either that:
\dcircle{1} the method itself is documented; or
\dcircle{2} in one of its parent classes/interfaces is documented.
For each method in the so derived dataset, we extracted \dcircle{1} its source code; and \dcircle{2} its documentation.
Eventually, our dataset consists of \num{46034} methods from the Android framework. 

\noindent
\textbf{Source Code Representation:}
Source code must be preprocessed as a piece of textual information before it can serve as input for machine-learning algorithms. In our work, we rely on \ctov~\cite{10.1145/3290353} (also see ection~\ref{sec:background}). Since the samples have different sizes, they must be transformed  into fixed-size numerical vectors. \ctov relies on a neural network that needs to be trained in order to generate source code vectors.
As the source code of the Android framework is Java code and the original pre-trained model available in the \ctov project repository\footnote{\url{https://github.com/tech-srl/code2vec}} has been trained on Java source code as well, we could have used this model. However, since Android code contains platform-specific semantic tokens that cannot be found in regular Java source code (e.g., Activity, BroadcastReceiver, etc.), we decided against this approach. Instead, we trained the  model with the source code from the Android framework to assure that our model properly captures the platform-specific tokens prevalent in Android.

After training the \ctov model with Android framework data, we fed the model with the \num{46034} Android methods previously extracted to generate their numerical value vectors. Eventually, \num{46034} vectors of size 384 were generated.

\noindent
\textbf{Documentation representation:}
In the same way as the source code, the documentation has to be embedded into fixed-sized numerical value vectors to be fed into machine learning algorithms. We relied on \sbert~\cite{reimers2019sentence} to generate those vectors. We leave experimentation with other models such as BERT~\cite{devlin2018bert} or RoBERTa~\cite{10.1007/978-3-030-84186-7_31} to future work. 

\begin{figure}
    \centering
    \includegraphics[scale=.6]{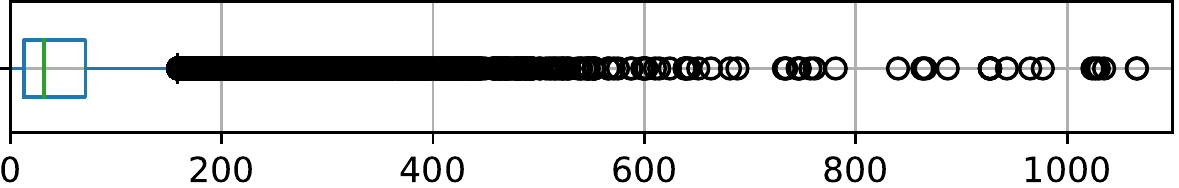}
    \caption{Distribution of the number of words in the documentation collected per Android method}
    \label{fig:doc_length}
\end{figure}

We used \sbert with the "paraphrase-mpnet-base-v2" pre-trained model, which is the one achieving the best performances\footnote{\url{https://www.sbert.net/docs/pretrained_models.html}} at the time of writing.
Eventually, all documentation of the \num{46034} previously gathered methods were converted into 768-value long vectors.
The distribution of the number of words in the documentation collected is available in Figure~\ref{fig:doc_length}.
Note that on average, the number of words in the documentation collected is 56, and the median is 32.
In future work, we will investigate the effect of text summarizing, i.e., learning from more compact texts.

\subsection{Deep Learning Architecture}
\label{sec:architecture}

\begin{figure}
    \centering
    \begin{adjustbox}{width=\columnwidth,center}
    \includegraphics[width=\columnwidth]{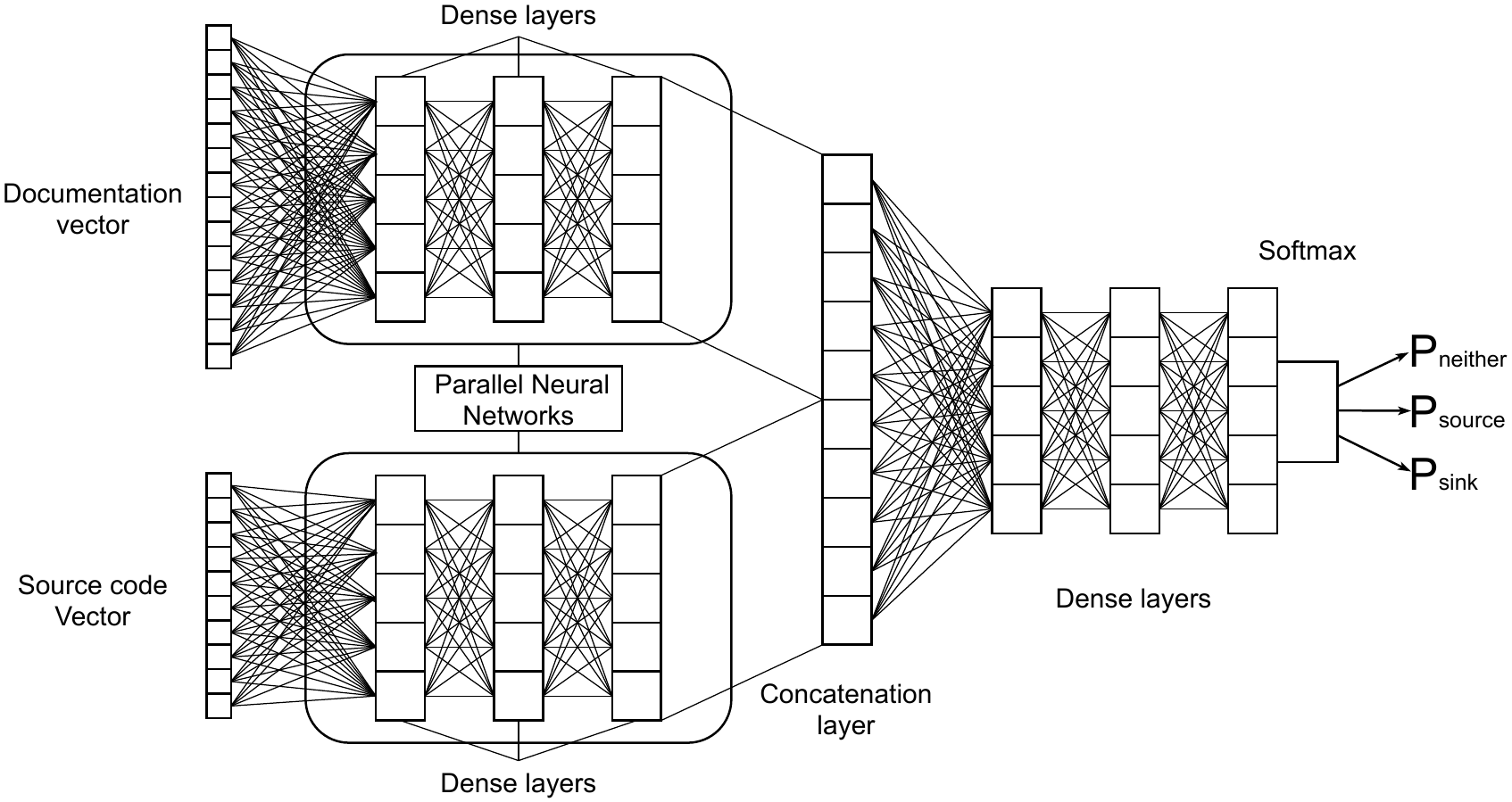}
    \end{adjustbox}
    \caption{\tool neural network architecture. 
    }
    \label{fig:network}
\end{figure}

Our deep learning model architecture is straightforward and aims at combining documentation and source code vectors into a single representation.
The overall architecture is available in Figure~\ref{fig:network}.
Since we are working on two different inputs (i.e., the documentation and the source code) of two different sizes, we decided to rely on two parallel and identical sub-neural networks and combine their output into a single vector that, in turn, is used for a classification task.
Each of those two parallel networks is built using a stack of three dense~\cite{huang2017densely} layers with ReLU~\cite{banerjee2019empirical} as activation function. 
They are used for extracting fixed-size features from the two inputs.
Thus, the first and the second sub-networks take as input the 768-long documentation vector and the 384-long source code vector, respectively, and provide as output two vectors of size 128 each.
Those outputs are combined using a concatenation layer that produces an unique 256-long vector.
We use this unique vector for a classification task carried out in 3 additional dense layers. A softmax~\cite{sharma2017activation} loss function is considered in the last dense layer in order to perform a multi-class classification, resulting in a classification as \source, \sink, or \neither.

\section{Evaluation}
\label{sec:evaluation}

To evaluate \tool, we address the following research questions:

\begin{description}
    \item[RQ1:] Do documentation and source code features provide complementary input for classification?
    \item[RQ2:] How does \tool perform in 10-fold cross-validation and how does it compare with \susi?
    \item[RQ3:] Can \tool make better predictions than \susi in the wild, i.e., with unseen data?
    \item[RQ4:] How does \tool perform on previously unseen methods?
    \item[RQ5:] How do the source and sink lists created by \tool and by \susi compare in data flow analysis?
\end{description}

\subsection{RQ1: Features complementarity}
\label{sec:eval:rq1}

\noindent
\textbf{Objective:}
In this section, we aim at evaluating to what extent both the source code and the documentation are needed to predict sensitive \source and \sink methods.
Intuitively, the source code and the documentation should contribute complementary pieces of semantic information.

\noindent
\textbf{Experimental Setup:}
To experimentally evaluate this hypothesis, we run and compare 4 configurations of \tool:

\begin{enumerate}
    \item Binary classification with \source and $\neg$\source
    \begin{enumerate}
        \item Only documentation
        \item Only source code
    \end{enumerate}
    \item Binary classification with \sink and $\neg$\sink
    \begin{enumerate}
        \item Only documentation
        \item Only source code
    \end{enumerate}
\end{enumerate}

\begin{table*}[ht!]
    \begin{adjustbox}{width=.9\textwidth,center}
        \begin{tabular}{|l|c|c|c|c|c|c|c|c|c|c||c|c|c|c|c|c|c|c|c|c|}
            \hline
            & \multicolumn{10}{c||}{\source prediction} & \multicolumn{10}{c|}{\sink prediction} \\
            \hline
            & \multicolumn{5}{c|}{Code} & \multicolumn{5}{c||}{Documentation} & \multicolumn{5}{c|}{Code} & \multicolumn{5}{c|}{Documentation}\\
            \hline
            & A & P & R & F & K & A & P & R & F & K & A & P & R & F & K & A & P & R & F & K \\
            \hline
            XGB &  0.86 & 0.74 & 0.62 &  0.67 &  0.58 & \bf 0.91 & \bf 0.87 & \bf 0.72 & \bf 0.78 & \bf 0.70 &  0.94 & 0.91 & 0.58 & 0.70 & 0.66 &  \bf 0.95 & 0.91 & \bf 0.67 & \bf 0.76 & \bf 0.68 \\
            \hline
            SVC & 0.82 & 0.61 & 0.64 & 0.62 & 0.49 & \bf 0.90 & \bf 0.75 & \bf  0.86 & \bf  0.80 & \bf  0.72 & 0.88 & 0.55 & 0.69 & 0.60 & 0.54 & \bf 0.94 & \bf 0.75 &  \bf 0.83 &  \bf 0.78 &  \bf 0.73 \\
            \hline
            DT & 0.80 & 0.56 & 0.60 & 0.57 & 0.49 & \bf 0.85 & \bf 0.66 & \bf 0.66 & \bf 0.66 & \bf 0.52 & \bf 0.88 &\bf  0.53 & \bf 0.54 & \bf 0.53 & 0.46 & 0.87 & 0.50 & 0.50 & 0.48 & \bf 0.50 \\
            \hline
            RF & 0.85 & 0.74 & \bf 0.56 & 0.62 &  0.58 & \bf 0.88 & \bf  0.92 & 0.55 & \bf 0.68 & \bf 0.58 & \bf 0.93 & \bf 0.94 &\bf  0.51 & \bf 0.65 & \bf 0.57 & 0.92 &  0.93 & 0.46 & 0.60 & 0.51 \\
            \hline
            GNB & 0.81 & 0.55 &  0.82 & 0.65 & 0.53 &\bf  0.87 &\bf  0.66 &\bf  0.85 & \bf 0.74 & \bf 0.65 & 0.82 & 0.40 &  \bf 0.77 & 0.52 & 0.42 & \bf 0.88 & \bf 0.54 & 0.76 & \bf 0.62 & \bf 0.56 \\
            \hline
            SGD & 0.82 & 0.61 & 0.57 & 0.58 & 0.50 &  \bf 0.91 & \bf 0.83 &\bf  0.79 &  \bf 0.80 &\bf  0.70 & 0.90 & 0.60 & 0.65 & 0.61 & 0.52 & \bf 0.93 & \bf 0.75 & \bf 0.74 &\bf  0.73 & \bf 0.70 \\
            \hline
            KNN & 0.85 & 0.73 & 0.56 & 0.63 & 0.56 & \bf 0.89 &\bf  0.82 & \bf 0.68 & \bf 0.74 & \bf 0.67 & 0.91 & 0.66 & 0.64 & 0.64 & 0.60 & \bf 0.92 & \bf 0.67 & \bf 0.82 &\bf  0.73 &\bf  0.69 \\
            \hline
            BDT1 & 0.85 & 0.74 & 0.50 & 0.59 & 0.51 & \bf 0.88 & \bf 0.86 & \bf 0.58 & \bf 0.68 &\bf  0.58 & \bf 0.92 &  \bf 0.84 &  \bf 0.49 & \bf 0.61 & \bf 0.54 & 0.91 & 0.73 & 0.45 & 0.55 & 0.49 \\
            \hline
            BDT2 & 0.85 & 0.74 & 0.50 & 0.59 & 0.51 & \bf 0.88 & \bf 0.86 & \bf 0.58 &\bf  0.68 &\bf  0.58 & \bf 0.92 & \bf 0.84 & \bf 0.49 & \bf 0.61 & \bf 0.54 & 0.91 & 0.73 & 0.45 & 0.55 & 0.49 \\
            \hline
            ET &  0.86 &  0.75 & \bf 0.56 & 0.63 &  \bf 0.58 & \bf 0.87 & \bf 0.89 & 0.51 & \bf 0.64 & 0.54 &  \bf 0.94 &  \bf 0.95 & \bf 0.60 &  \bf 0.73 &  \bf 0.67 & 0.93 & 0.93 & 0.49 & 0.62 & 0.57 \\
            \hline
            ADA1 & 0.84 & 0.66 & 0.62 & 0.63 & 0.50 & \bf 0.89 & \bf 0.78 & \bf 0.70 & \bf 0.73 & \bf 0.70 & 0.91 & 0.74 & 0.58 & 0.63 & 0.55 & \bf 0.94 & \bf \bf 0.78 &\bf  0.73 & \bf 0.74 &\bf  0.70 \\
            \hline
            ADA2 & 0.84 & 0.66 & 0.62 & 0.63 & 0.50 & \bf 0.89 & \bf 0.78 & \bf 0.70 & \bf 0.73 & \bf 0.70 & 0.91 & 0.74 & 0.58 & 0.63 & 0.55 &\bf  0.94 & \bf \bf 0.78 & \bf 0.73 & \bf 0.74 & \bf 0.70 \\
            \hline
            GB &  0.86 & 0.72 & 0.62 & 0.66 & 0.57 & \bf 0.90 & \bf 0.83 & \bf 0.68 & \bf 0.75 & \bf 0.69 & \bf 0.93 & \bf 0.90 & 0.52 & 0.64 & 0.59 & 0.92 & 0.85 & \bf 0.53 & 0.64 & 0.59 \\
            \hline
            NN & 0.84 & 0.69 & 0.56 & 0.60 & 0.50 & \bf 0.90 & \bf 0.86 & \bf 0.68 & \bf 0.75 & \bf 0.64 & \bf 0.90 &\bf  0.77 &\bf  0.34 &\bf  0.45 &\bf  0.33 & 0.87 & 0.00 & 0.00 & 0.00 & 0.00 \\
            \hline
        \end{tabular}
    \end{adjustbox}
    \caption{Results of binary classification on multiple classifiers with code and documentation. (A = Accuracy, P = Precision, R = Recall, F = F$_1$ score, K = Kappa score)}
    \label{table:binary_results}
\end{table*}

Note that we tested these configurations on multiple classifiers, i.e., we exchanged the dense classification layer in Figure~\ref{fig:network} with other classifiers. We did so to ensure that RQ1 is answered in-depth and not dependent on a single classification approach. For each binary classification described above, we performed a stratified 10-fold cross-validation~\cite{kohavi1995study,10.2307/2984809}, and for each iteration we retained the methods that were mis-classified considering only the documentation and well-classified using only the source code, and vice-versa. The results using only code resp. only documentation are available in Table~\ref{table:binary_results}.
We notice that predictions using only documentation yield better results both for source and sink predictions.

\begin{figure}[ht!]
    \centering
    \begin{subfigure}[b]{0.4\columnwidth}
         \centering
         \includegraphics[width=\textwidth]{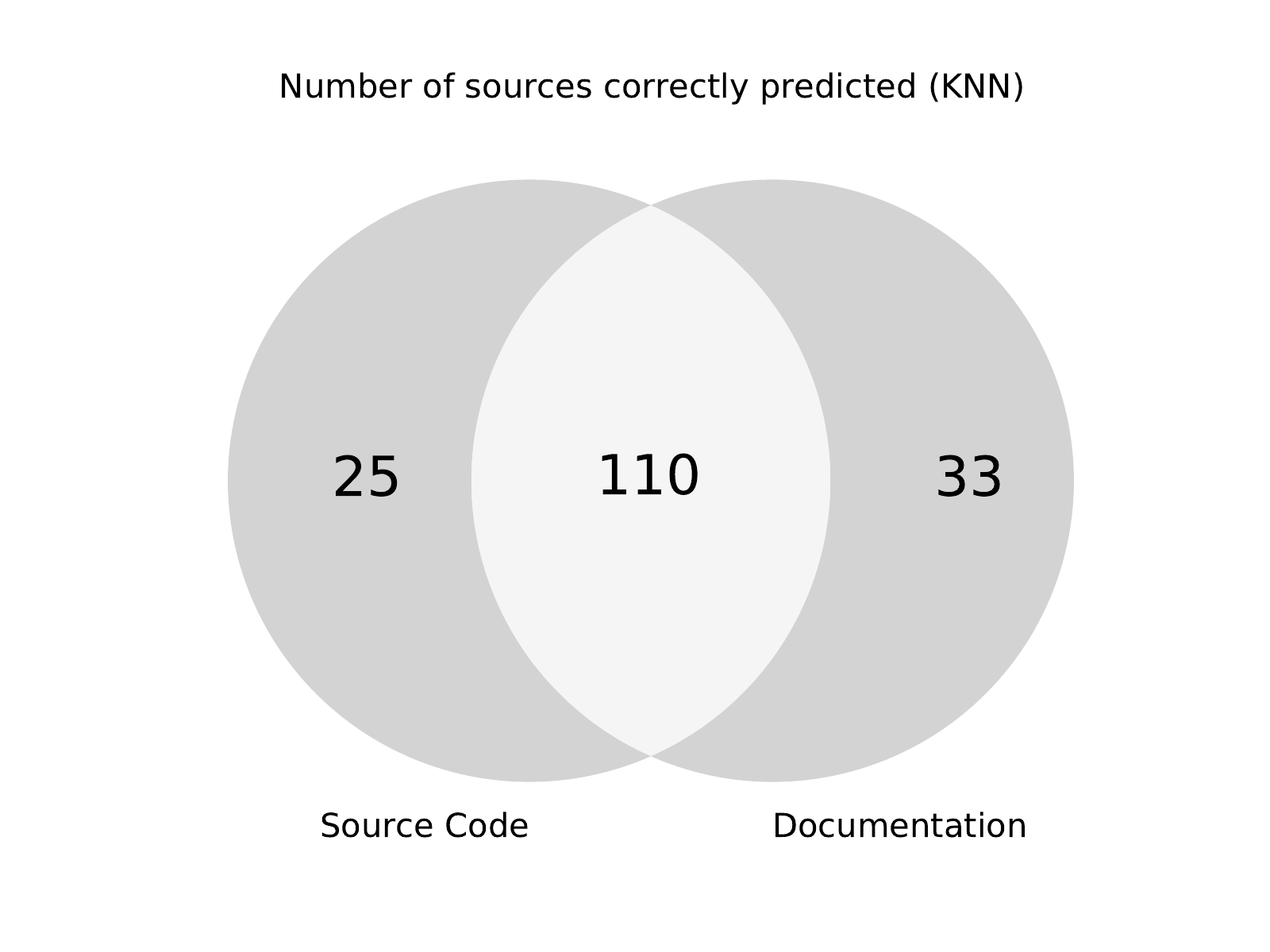}
         \caption{Sources prediction}
         \label{fig:rq1:sources}
     \end{subfigure}
     \begin{subfigure}[b]{0.4\columnwidth}
         \centering
         \includegraphics[width=\textwidth]{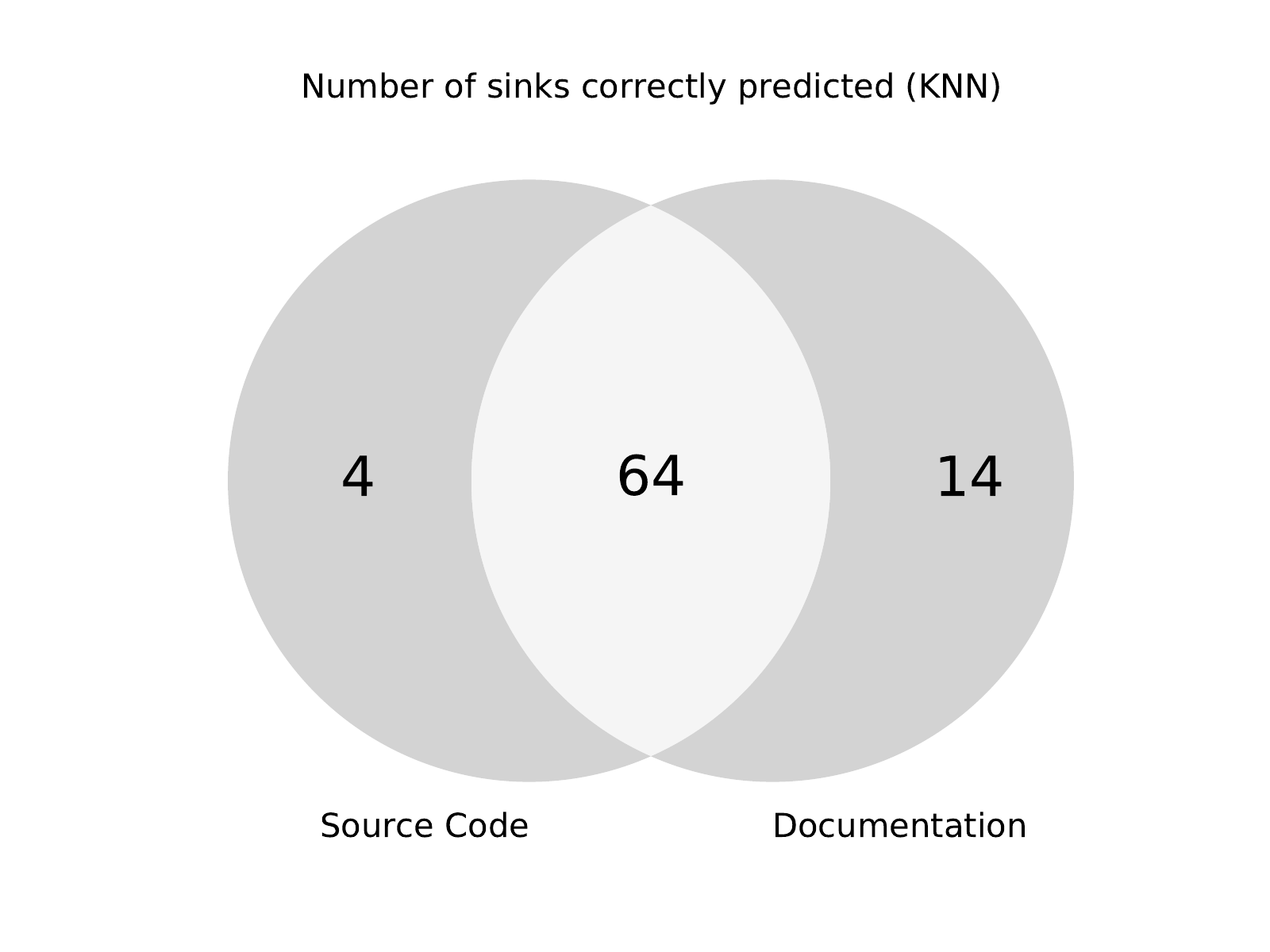}
         \caption{Sinks prediction}
         \label{fig:rq1:sinks}
     \end{subfigure}
    \caption{Number of sources and sinks correctly predicted with the KNN algorithm.}
\end{figure}

Further, we computed the overall number of methods that were mis-classified with the documentation but well-classified with the source code, and vice-versa for sources and sinks and for each classifier. Figures~\ref{fig:rq1:sources} and~\ref{fig:rq1:sinks} illustrate the results of this experiment for the KNN algorithm. 
There we can see that classification with the source code can correctly predict 25 sources that classification with documentation cannot, and 33 vice-versa.
In the same way, classification with the source code can correctly predict 4 sinks that classification with documentation cannot, and 14 vice-versa.
This shows that, although classification with documentation is better, it misses some samples that classification with source code does not.
This shows the need to use both features for our work to improve our model capabilities.

\highlight{
\textbf{RQ1 answer:}
Both the documentation and the code independently bring additional information with regard to the other, i.e., several sources/sinks could only be found with the documentation, or only with code features.
}

\subsection{RQ2: 10-fold cross validation on \tool's model and comparison with the state-of-the-art \susi}
\label{sec:eval:rq2}

\noindent
\textbf{Objective:}
In this section, we investigate whether our approach can better identify sensitive \source and \sink methods than the state-of-the-art \susi approach. We use \tool with source code and documentation as input. Note that classifying \source and \sink methods is not a binary classification problem since a method can either be: \dcircle{1} a \source; \dcircle{2} a \sink; or \dcircle{3} \neither. Hence, our model relies on a multiclass classification.

\noindent
\textbf{\tool performances:} To evaluate our approach, we apply a stratified 10-fold cross-validation and compute the following metrics: \dcircle{1} precision ($\frac{|TP|}{|TP| + |FP|}$); \dcircle{2} recall ($\frac{|TP|}{|TP| + |FN|}$); and \dcircle{3} F$_1$ score ($2 \times \frac{precision \times recall}{precision + recall}$), with $TP$ = True Positive, $FP$ = False Positive, and $FN$ = False Negative.

In Table~\ref{table:tool_results}, we present the results of \tool.
Note that we retain the activation function that yielded the best results for our neural network, i.e., ReLU~\cite{agarap2018deep} (we tested the following activation functions: ReLU, sigmoid, tangent, and several combinations of these functions).

\begin{table}
    \begin{adjustbox}{width=.7\columnwidth,center}
        \begin{tabular}{lrrr}
            \Xhline{2\arrayrulewidth}
            & Precision & Recall & F$_1$ score  \\
            \hline
            \source & 0.82 & 0.88 & 0.85 \\
            \sink & 0.93 & 0.87 & 0.90 \\
            \neither & 0.95 & 0.93 & 0.94 \\
            \hline
            Macro Average & 0.90 & 0.89 & 0.89 \\
            Weighted Average & \textbf{0.91} & \textbf{0.91} & \textbf{0.91} \\
            \Xhline{2\arrayrulewidth}
        \end{tabular}
    \end{adjustbox}
    \caption{\tool performances}
    \label{table:tool_results}
\end{table}

\begin{table}
    \begin{adjustbox}{width=.7\columnwidth,center}
        \begin{tabular}{lrrr}
            \Xhline{2\arrayrulewidth}
            & Precision & Recall & F$_1$ score  \\
            \hline
            \source & 0.83 & 0.85 & 0.84  \\
            \sink & 0.86 & 0.71 & 0.78 \\
            \neither & 0.89 & 0.91 & 0.90 \\
            \hline
            Macro Average & 0.86 & 0.82 & 0.84 \\
            Weighted Average & \textbf{0.87} & \textbf{0.87} & \textbf{0.87} \\
            \Xhline{2\arrayrulewidth}
        \end{tabular}
    \end{adjustbox}
    \caption{\susi performances with our ground-truth}
    \label{table:susi_results}
\end{table}

\noindent
\textbf{Comparison with \susi:} \susi is not intended to generate \emph{privacy-sensitive} \source and \sink methods, as explained in Section~\ref{sec:introduction}. Therefore, a direct comparison between our approach \tool and the pre-trained \susi would be unfair~\cite{10.5555/1248547.1248548}. We therefore trained \susi on our own ground truth to evaluate it and compare it with \tool.

Table~\ref{table:susi_results} shows the results of a 10-fold cross validation. First, we notice that both \tool and \susi independently yield better results for the \neither class compared to \source and \sink classes. This is expected since the training data set is highly imbalanced towards \neither methods -- most of the Android API is not a source or sink. While this imbalance may be relieved using over-/undersampling techniques or class weights, it does not affect the comparative performance of the tools.

Second, \tool clearly yields slightly better performance than \susi to classify sensitive \source and \sink methods. \tool outperforms \susi by 4 points for the precision, the recall, and the F$_1$ score.
\highlight{
\textbf{RQ2 answer:}
The performance of our approach \tool achieves an F$_1$ score of 91\% to identify \source and \sink methods in the Android framework.
Furthermore, on the same training set, \tool achieves a slightly better score than the state-of-the-art \source and \sink classifier \susi. 
}

\subsection{RQ3: \tool and \susi comparison in the wild}
\label{sec:eval:rq3}

\noindent
\textbf{Objective:}
This research question aims to compare which methods are identified as sensitive \source or \sink methods by \tool, and which ones by \susi. 
To do so, we check the (non-)overlap to judge the performance of both approaches outside of the 10-fold cross-validation. We apply \susi and \tool on the full Android SDK, containing mostly non-labeled methods.

\noindent
\textbf{Experimental setup:}
To compare \tool against \susi, we generate the following lists of \source and \sink methods (sizes shown in Table~\ref{table:number_methods}):
\begin{enumerate}
    \item \textbf{\tool}: the lists generated by \tool on Android 30 (API version 11), trained on our ground truth.
    \item \textbf{\susiOriginal}: the lists generated by \susi on a more recent version of Android (i.e., 30), trained on \susi's original ground truth.
    \item \textbf{\susiNew}: the lists generated by \susi on a more recent version of Android (i.e., 30), trained on our ground truth.
\end{enumerate}

\begin{table}
    \begin{adjustbox}{width=.7\columnwidth,center}
        \begin{tabular}{lrr}
            \Xhline{2\arrayrulewidth}
            Lists & \# \source & \# \sink \\
            \Xhline{2\arrayrulewidth}
            \tool & \num{15105} & \num{1061} \\
            \hline
            \susiOriginal & \num{25369} & \num{5913} \\
            \hline
            \susiNew & \num{12082} & \num{1010} \\
            \Xhline{2\arrayrulewidth}
        \end{tabular}
    \end{adjustbox}
    \caption{Number of \source and \sink methods in the lists generated by \tool and \susi}
    \label{table:number_methods}
\end{table}

\begin{figure}[ht!]
    \centering
    \begin{subfigure}[b]{0.4\columnwidth}
         \centering
         \includegraphics[width=\textwidth]{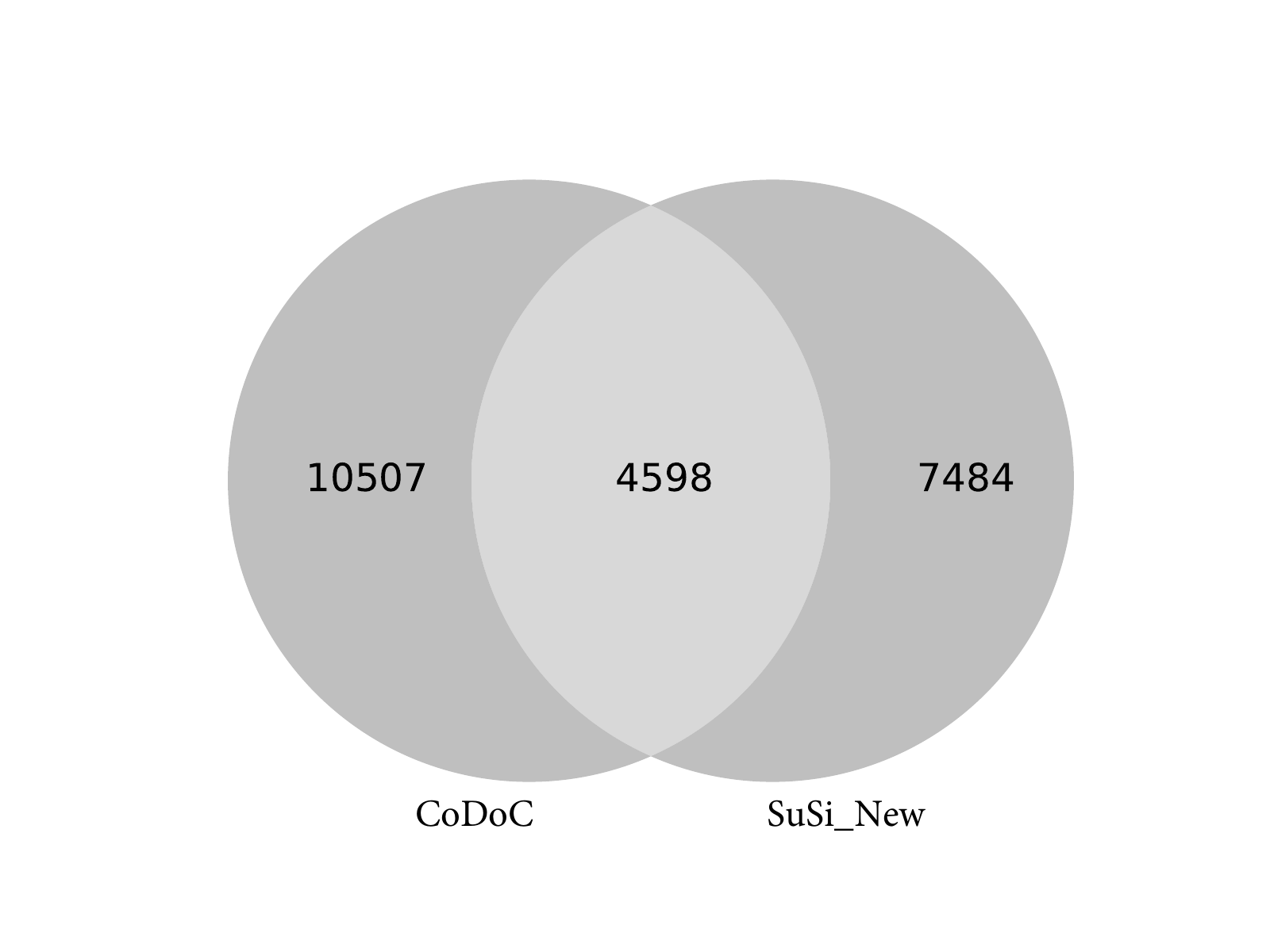}
         \caption{Sources prediction}
         \label{fig:rq4:sources_new}
     \end{subfigure}
     \begin{subfigure}[b]{0.4\columnwidth}
         \centering
         \includegraphics[width=\textwidth]{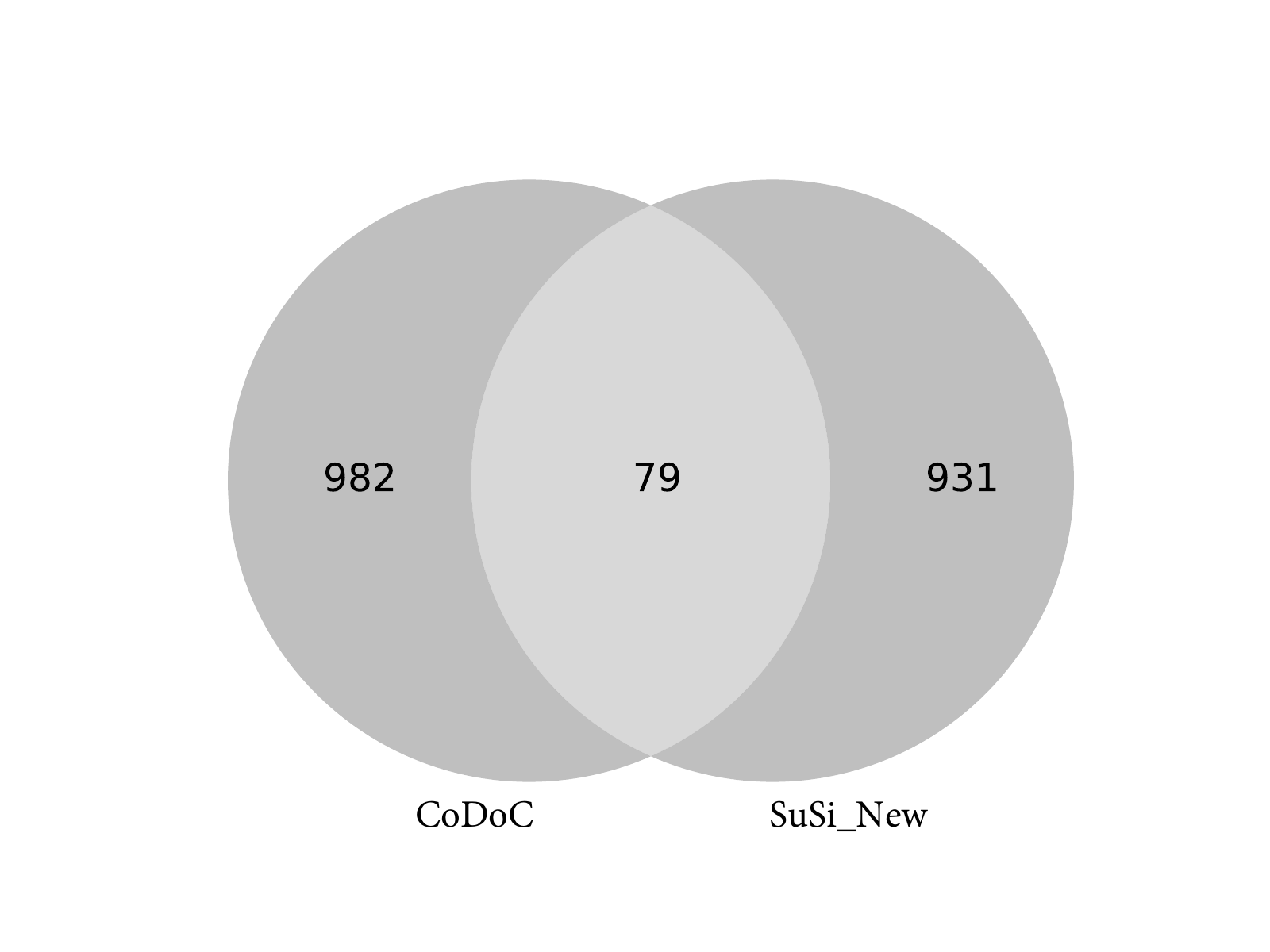}
         \caption{Sinks prediction}
         \label{fig:rq4:sinks_new}
     \end{subfigure}
     \begin{subfigure}[b]{0.4\columnwidth}
         \centering
         \includegraphics[width=\textwidth]{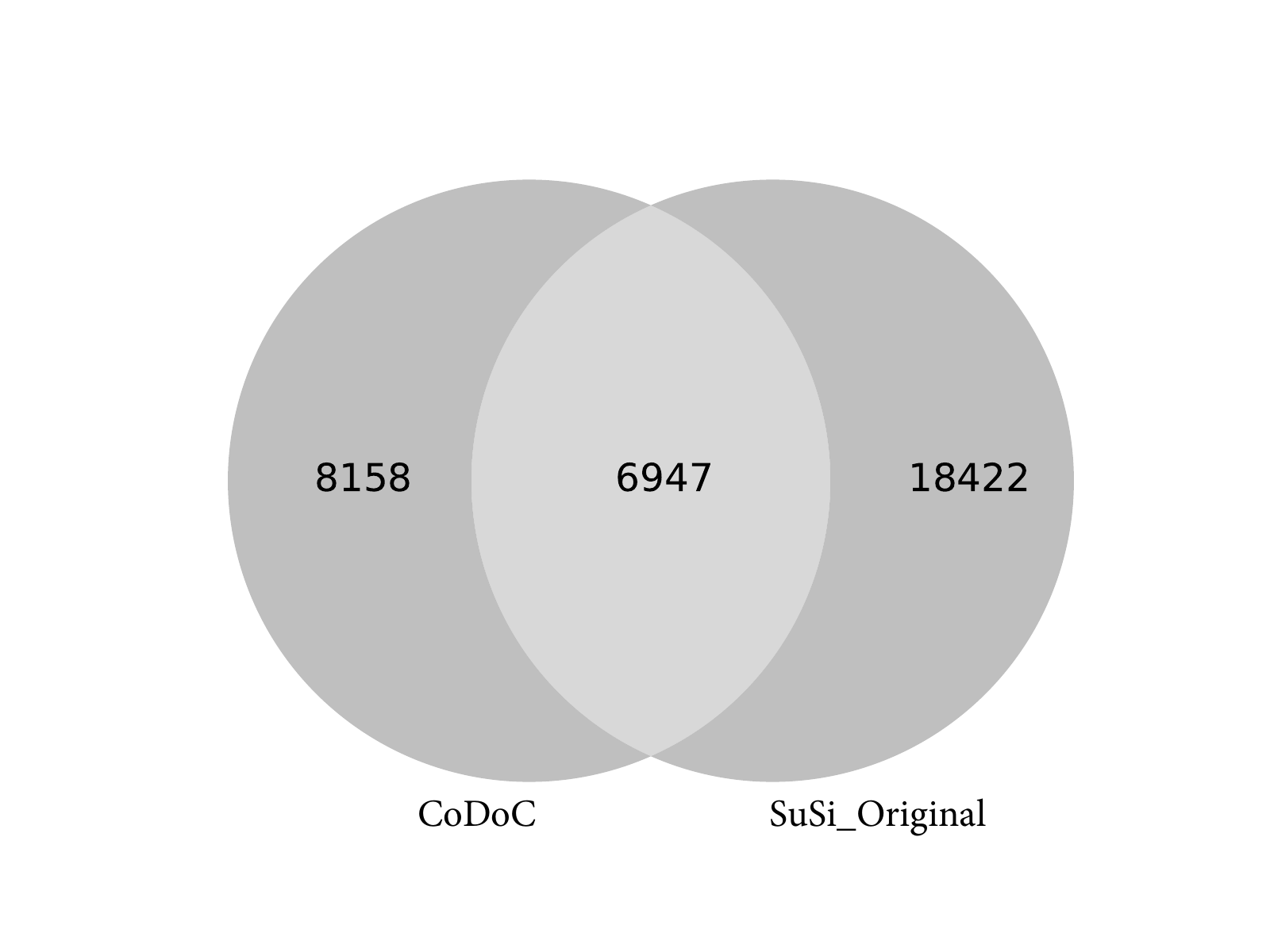}
         \caption{Sources prediction}
         \label{fig:rq4:sources_original}
     \end{subfigure}
     \begin{subfigure}[b]{0.4\columnwidth}
         \centering
         \includegraphics[width=\textwidth]{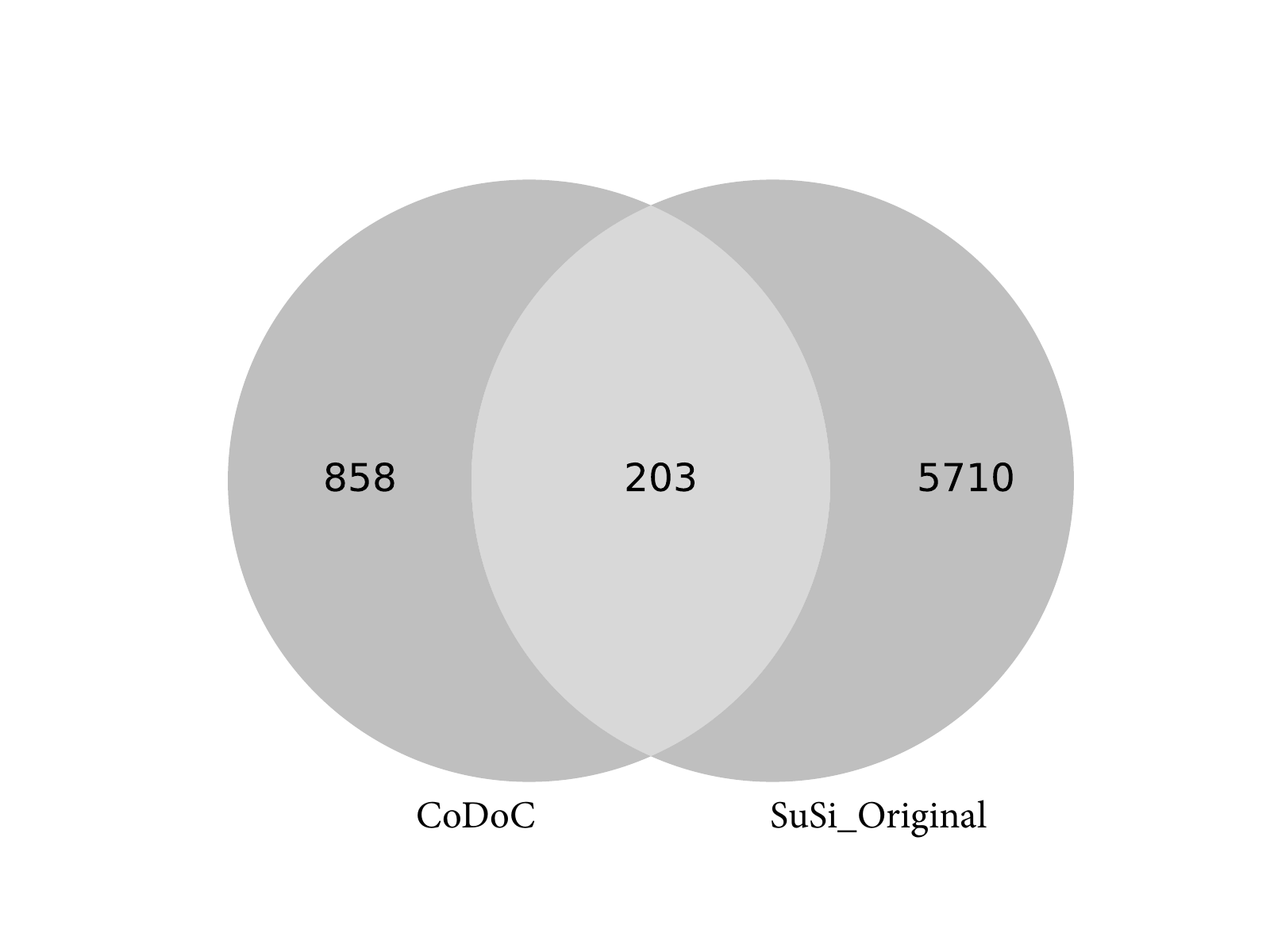}
         \caption{Sinks prediction}
         \label{fig:rq4:sinks_original}
     \end{subfigure}
    \caption{Overlap between \source and \sink methods lists of \susi and \tool.}
    \label{fig:overlap}
\end{figure}

We first analyze the overlap between \susi's and \tool's lists in Figure~\ref{fig:overlap}. 
We notice that for both \susiOriginal and \susiNew, and both \source and \sink methods, \tool and \susi do not have much in common.

We manually curated 6 data sets for further inspection:
\begin{enumerate}
    \item We select 100 non-sensitive \source methods classified by \susiOriginal and compare \tool's predictions
    \item We select 100 misclassified \sink methods predicted by \susiOriginal and compare \tool's predictions
    \item We select 100 non-sensitive \source methods classified by \susiNew and compare \tool's predictions
    \item We select 100 misclassified \sink methods predicted by \susiNew and compare \tool's predictions
    \item We select 100 misclassified \source methods predicted by \tool and compare with both \susiOriginal and \susiNew
    \item We select 100 misclassified \sink methods predicted by \tool and compare with both \susiOriginal and \susiNew
\end{enumerate}

\noindent
\textbf{Methods selection:}
Regarding the sources, the authors randomly browsed the \source methods yielded by \susiOriginal, \susiNew, and \tool, and consulted the source code and the documentation to ensure the sensitiveness of the methods.
In case the methods did not return sensitive values, they were retained for this research question. 
The authors stopped after 100 \source methods (which is statistically significant at a 95\% confidence level and a confidence interval $\pm$ 10\% from the \num{25369} sources of \susiOriginal, the \num{12082} of \susiNew, and the \num{15105} of \tool). 
Regarding the sinks, the same procedure was applied as for sources, except the authors checked if the methods could send data out of the app (100 sinks are statistically significant at a 95\% confidence level, and a confidence interval $\pm$ 10\% from the \num{5913} sinks of \susiOriginal, the \num{1010} of \susiNew, and the \num{1061} of \tool).
An example of a non-sensitive \source method misclassified is: "android.bluetooth.BluetoothCodecStatus.describeContents()" that can be seen in Listing~\ref{code:non_sensitive_source}.
Indeed, the documentation and the source code are very explicit: this method only returns 0.
An example of a \sink method misclassified is: "com.android.inter\-nal.util.FastMath.round\-(float)", the source code and documentation of which are available in Listing~\ref{code:non_sink}.
Indeed, this method is only intended to round a float number. It does not write any value outside the app.

Eventually, we gathered 6 datasets with 100 methods each.

\begin{listing}
\begin{minted}{java}
/**
 * Always returns 0
 *
 * @return 0
 * @hide
 */
public int describeContents() {
    return 0;
}
\end{minted}
    \caption{Source code of method "\texttt{describeContents}" of class "\texttt{android.bluetooth.BluetoothCodecStatus}"}
    \label{code:non_sensitive_source}
\end{listing}

\begin{listing}
\begin{minted}{java}
/**
 * Fast round from float to int. This is faster than Math.round()
 * thought it may return slightly different results. It does not try to
 * handle (in any meaningful way) NaN or infinities.
 */
public static int round(float value) {
    long lx = (long) (value * (65536 * 256f));
    return (int) ((lx + 0x800000) >> 24);
}
\end{minted}
    \caption{Source code of method "\texttt{round}" of class "\texttt{com.android.internal.util.FastMath}"}
    \label{code:non_sink}
\end{listing}

\noindent
\textbf{Results:}
The overall results are available in Table~\ref{table:misclassified_methods}.
From the 100 non-sensitive \source methods classified as sources by \susi, \tool only classified 13 as sensitive \source methods.
From the 100 non-\sink methods classified as sinks by \susi, \tool only classified 4 as \sink methods.

\begin{table}
    \begin{adjustbox}{width=.8\columnwidth,center}
        \begin{tabular}{l|cc|cc}
            \Xhline{2\arrayrulewidth}
            Prediction & \susiOriginal & \tool & \susiNew & \tool\\
            \Xhline{2\arrayrulewidth}
            \source & 100 & 13 & 100 & 26 \\
            \hline
            \sink & 100 & 4 & 100 & 18 \\
            \Xhline{2\arrayrulewidth}
        \end{tabular}
    \end{adjustbox}
    \caption{\tool performance on \susi's misclassified \source and \sink methods}
    \label{table:misclassified_methods}
\end{table}

\begin{table}
    \begin{adjustbox}{width=.8\columnwidth,center}
        \begin{tabular}{l|cc|cc}
            \Xhline{2\arrayrulewidth}
            Prediction & \tool & \susiOriginal & \tool & \susiNew\\
            \Xhline{2\arrayrulewidth}
            \source & 100 & 32 & 100 & 6 \\
            \hline
            \sink & 100 & 3 & 100 & 0 \\
            \Xhline{2\arrayrulewidth}
        \end{tabular}
    \end{adjustbox}
    \caption{\susi performance on \tool's misclassified \source and \sink methods}
    \label{table:misclassified_methods_codoc}
\end{table}

\noindent
\textbf{Discussion:}
These results, in the wild, do not confirm that \tool is better than \susi (contrary to what is shown in Section~\ref{sec:eval:rq2}), nor the other way around, due to the very low overlap between the \source and \sink methods predicted by both tools. Tables~\ref{table:misclassified_methods} and~\ref{table:misclassified_methods_codoc} do not show that any approach is stronger.
Better, taken independently, Table~\ref{table:misclassified_methods} shows that \tool is preferable, while Table~\ref{table:misclassified_methods_codoc} shows that \susi is preferable.
Hence, even if intuitively, by relying on both source code and documentation, better results are expected, in practice, it is not true according to our experiments and observations. 

\highlight{
\textbf{RQ3 answer:}
While the outputs of \tool and \susi differ, neither is strictly superior over the other. Though \tool excels on \susi's misclassified results, the inverse is also correct. 
}

\subsection{RQ4: Real-world performance of \tool}
\label{sec:eval:rq4}

\noindent
\textbf{Objective:}
In this section, we aim to qualitatively evaluate \tool's predicted lists of \source and \sink methods, which is the main goal of this approach, i.e., \emph{generating lists of \source and \sink methods actionable to other tools, e.g., data leak detectors}.
To do so, we randomly selected 100 \source and 100 \sink methods in the lists generated by \textbf{\tool}, which is statistically significant for the dataset at a 95\% confidence level and a confidence interval $\pm$ 10\%, and inspected each method based on two criteria: \dcircle{1} the sensitiveness of \source methods for user privacy; and \dcircle{2} the fact that \sink methods can make data leave the app space.

\noindent
\textbf{Results:}
The results of our manual analyses are available in Table~\ref{table:fp_tp_measurement}.
Although, as seen in Section~\ref{sec:eval:rq2}, \tool achieves a score of 82\% precision in the lab to classify \source methods, in the wild, it reaches a false positive rate of 90\%.
Same conclusion for \sink methods for which though \tool achieves a precision score of 93\% to classify \sink methods, in the wild, it reaches a false positive rate of 56\%.

\begin{table}
    \begin{adjustbox}{width=.7\columnwidth,center}
        \begin{tabular}{l|cc}
            \Xhline{2\arrayrulewidth}
             & False positives & True positives\\
            \Xhline{2\arrayrulewidth}
            \source & 90 & 10 \\
            \hline
            \sink & 56 & 44 \\
            \Xhline{2\arrayrulewidth}
        \end{tabular}
    \end{adjustbox}
    \caption{Source and sink methods' false and true positives' rates predicted by \tool}
    \label{table:fp_tp_measurement}
\end{table}

\highlight{
\textbf{RQ4 answer:}
Results indicate that though \tool achieves high-performance scores when assessing its underlying deep learning model, in the wild, it achieves poorly with over 90\% false positive for \source methods and 56\% false positive for \sink methods.
 }

\subsection{RQ5: False positive measurement in sensitive data leak detection}
\label{sec:eval:rq5}

To evaluate the effect of the false positives generated by \tool on real-world applications, we utilize \flowdroid~\cite{arzt2014flowdroid} to find data leaks in Android apps. Intuitively, a false positive in a list of sources or sinks is only relevant if it leads to spurious leaks in the data flow analysis. A method that is never used in an app might be on a source or sink list, but does not have any negative effect in practice.

For this evaluation, we randomly selected 500 popular apps from the 2022 \gp across all available categories. 
For each app analysis, we set \flowdroid timeouts to 5 min for data flow analysis and 3 min for callback collection and configured the JVM with 768GB maximum heap size.
We ran the analysis on a system with 144 logical cores backed by four physical Intel Xeon Gold 6254 CPUs.
Note that we focus on the quality of sources and sinks and not the performance of the data flow analysis. 
We, therefore, opted for a system with sufficient resources to scale to large apps.

We configure \flowdroid with three different lists:
\begin{enumerate}
    \item \tool: The list generated by \tool on Android 30.
    \item \susiNew: The list generated by \susi, where \susi was trained on the ground truth presented in this paper and classified methods of Android version 30.
    \item \susiOriginal: The list generated by \susi, where \susi was trained on the original \susi training data and classified methods of Android version 30.
\end{enumerate}

For each of these lists ($Src_{NN}, Snk_{NN}$ with $NN \in \{\susi, \susiNew, \tool\}$), \flowdroid yields a set of data flows $Flow_{NN}$. We then use the data flows $Flow_{NN}$ to remove all sources and sinks from the lists that are not used in at least one data flow. This leads to a reduced list of sources and sinks, i.e. $\widehat{Src}_{NN}, \widehat{Snk}_{NN}$. We validate these lists by hand and count the number of methods that lead to leaks, but that are not actually privacy-sensitive.

For \tool, we find \texttt{StringBuilder.toString()}, which is clearly a false positive, to be the most commonly used ``source'' in the data flow analysis (72\% of all flows). The second most common source was \texttt{StringBuffer.toString()} with around 8\% of all flows. The sinks are more reasonable, with the Android log methods being the most prevalent ones (21\% of all flows).

The \susiNew results lead to far fewer flows (3,410 instead of \num{71211} for \tool). The used sources and sinks are more widely distributed, i.e., the top source only accounts for 10\% of all sources. Still, the used sources and sinks are mostly false positives.

For \susiOriginal, we find the most flows (\num{153558}). The structure of used sources and sinks resembles \susiNew, i.e., a wide variety of methods, most of which are false positives.

\highlight{
\textbf{RQ5 answer:}
The results show that the false positives generated by \susi and \tool have a major negative impact on the precision of the data flow analysis that uses these lists of sources and sinks.
}
\section{Discussion}
\label{sec:discussion}

We designed our study and approach under the hypothesis that adding more semantic, i.e., using code and documentation together, would provide better results than the current state of the art to classify \source and \sink methods. Further, we integrated recent advances in machine learning. Unfortunately, empirical results show that \tool performs poorly in practice. More precisely, our investigations show that:

\begin{itemize}
    \item Although code and documentation are complementary to predict \source and \sink methods, \tool performs poorly in the wild.
    \item Although \tool achieves good lab results and better lab results than \susi, i.e., precision, recall, and F$_1$ score of 91\% compared to a precision, recall and F$_1$ score of 87\%, it performs poorly in the wild.
    \item \tool generates many false positives when applied to Android framework methods, i.e., classifies \source and \sink methods that are not \source and \sink methods.
    \item The false positives in the \source and \sink lists lead to false positives in the data flow analysis which renders the lists unfit for real-world data leak detection scenarios.
\end{itemize}
\section{Limitations and Threats to validity}
\label{sec:threat_to_validity}

\tool relies on two inputs to make its prediction, namely the source code and the documentation of Android methods. We acknowledge that the Android framework contains undocumented methods which cannot be taken into account by \tool. The lack of method documentation makes \tool miss some methods to classify, hence, sensitive \source and \sink methods are certainly missed. However, the proportion of methods missed is too low (i.e., \num{18.6}\%) to fully explain the poor real-world performance of the approach.

\tool relies on supervised machine learning techniques which, by definition, need labeled data. Therefore, we performed manual labeling based on our expertise to label Android framework methods as \source, \sink, or \neither. 
Consequently, though we observed a strict and consistent procedure, our labels can be influenced by human subjectivity.
Nonetheless, we make public all of our artifacts to the research community to mitigate this threat to validity.

Our training set is limited to \total samples across three classes. This might not be enough training data. We will explore the use of data augmentation in future work.

Sensitiveness is a concept not well-defined, especially for technical frameworks, and exposed to human subjectivity since there is no formal definition of what it is.
Also, the authors noticed during manual labeling that sometimes there is only a fine line between a sensitive value and a non-sensitive one.
Therefore, the choices regarding sensitiveness can be biased based on human subjectivity.

As already described and motivated in Section~\ref{sec:manual_labeling}, our manual labeling process was performed without taking into account \susi's \neither methods list.
However, we note that since \susi yields very good results on the \neither category~\cite{arzt2013susi}, there is a high chance that this list contained well-classified samples, hence being more representative than the ones misclassified as \source and \sink methods.
\section{Related work}
\label{sec:related_work}

In this section, we present the related works available in the literature that are close to our work.

\noindent
Taint analysis, which requires proper lists of sources and sinks,  is used for a variety of purposes: 
vulnerability detection~\cite{10.5555/1251398.1251416,Cai2016,newsome2005dynamic}, sensitive data leak detection~\cite{arzt2014flowdroid,li2015iccta,9402001,enck2014taintdroid,10.1145/2660267.2660357,gordon2015information,10.1145/2660267.2660357,9402001,6394931}, hidden behavior detection~\cite{zhao2020automatic,8342924}, malware detection~\cite{7976989} or bad practices detection~\cite{7972733}.
All of these works require lists of \source and \sink methods that have to be defined in advance. If these lists are not complete, the approaches may miss important data flows. Therefore, automated techniques were proposed to catch as many \source and \sink methods as possible, aiming to reach completeness. In 2012, Gibler \& al.~\cite{10.1007/978-3-642-30921-2_17} proposed an approach to automatically detect \source and \sink methods based on mappings between methods and the permissions needed to invoke those methods. Methods requiring sensitive permissions were considered sources. Methods requiring the \texttt{INTERNET} permission were considered sinks.
However, not all sensitive methods need permissions in the Android framework~\cite{10.1145/3292006.3300023}. Thus, permission-based approaches miss relevant sources and sinks.
Our work, on the other hand, considers all methods in the Android framework regardless of required permissions.

Two years later, Arzt \& al.~\cite{arzt2013susi} proposed \susi, an automated approach that relies on machine learning to classify \source and \sink methods in the Android framework.
\susi relies on features based on the method signature (e.g., the method name, the parameter types, the return value, the modifiers, etc.) and based on dataflow features.
In contrast to \susi, our approach \tool relies on a multi-input deep-learning classifier based on: \dcircle{1} the source code, and \dcircle{2} the documentation of a method.

More recently, Wongwiwatchai \& al.~\cite{9163043} proposed an approach to detect privacy leaks in Android apps.
The authors did not rely on existing lists of \source and \sink methods. Rather, they defined their own lists.
To do so, the authors study well-known frameworks on data protection (e.g., the GDPR) and constitute a list of personal information commonly defined in these regulations.
The process of mapping personal information (e.g., an age) to Android APIs is opaque in the paper. Hence it is difficult to judge the approach's comprehensiveness and the rate of false positives.
In contrast, our approach aims to automatically and systematically map sensitive API methods to sensitive data with machine learning techniques.
\section{Conclusion}
\label{sec:conclusion}

As described in Section~\ref{sec:eval:rq4}, \tool, likewise \susi, does not provide actionable results in the wild.
Indeed, although we have shown in Section~\ref{sec:eval:rq2} that \tool outperforms \susi with a score of 91\% on our ground truth, our manual evaluations have shown that \tool performs poorly on the Android framework methods.
Hence, the resulting \source and \sink methods' list produced cannot be relied upon in real-world data leak detection scenarios.
This negative result and the literature~\cite{8776652,JUNAID201692,8952502}  show: \dcircle{1} the problem of classifying \source and \sink methods is not trivial; and \dcircle{2} there is an urgent need of a community effort to produce an actionable list of \source and \sink methods for sensitive data leak detection in Android apps carrying highly sensitive data about end users.
\section{Data Availability}
\label{sec:data_availability}

For the sake of Open Science, we provide to the community all the artifacts used in our study.
In particular, we make available the datasets used, the source code of our prototype and the scripts to execute \tool, our manually labeled datasets, the vector representation of source code and documentation, and \susi related artifacts. The project's repository including all artifacts is available at: \url{https://github.com/JordanSamhi/CoDoC}
\section{Acknowledgment}
\label{sec:acknowledgment}
This research work has been funded by the German Federal Ministry of Education and Research and the Hessian Ministry of Higher Education, Research, Science and the Arts within their joint support of the National Research Center for Applied Cybersecurity ATHENE.
Additionally, this work was partly supported by the Luxembourg National Research Fund (FNR), under projects Reprocess C21/IS/16344458 and the AFR grant 14596679.

\clearpage

\bibliographystyle{plain}

\bibliography{bib}

\end{document}